# LinkGlide-S: A Wearable Multi-Contact Tactile Display Aimed at Rendering Object Softness at the Palm with Impedance Control in VR and Telemanipulation

Miguel Altamirano Cabrera, Jonathan Tirado, Juan Heredia, and Dzmitry Tsetserukou

*Abstract*— LinkGlide-S is a novel wearable hand-worn tactile display to deliver multi-contact and multi-modal stimuli at the user's palm. The array of inverted five-bar linkages generates three independent contact points to cover the whole palm area. The independent contact points generate various tactile patterns at the user's hand, providing multi-contact tactile feedback. An impedance control delivers the stiffness of objects according to different parameters. Three experiments were performed to evaluate the perception of patterns, investigate the realistic perception of object interaction in Virtual Reality, and assess the users' softness perception by the impedance control. The experimental results revealed a high recognition rate for the generated patterns. These results confirm that the performance of LinkGlide-S is adequate to detect and manipulate virtual objects with different stiffness. This novel haptic device can potentially achieve a highly immersive VR experience and more interactive applications during telemanipulation.

## I. INTRODUCTION

Nowadays, teleoperation and telepresence represent one of the most natural applications of haptics, while the most effective combination of various haptic feedbacks remain to be the subject of active research. Many Virtual Reality (VR) applications have been introduced, such as simulators and games, and are becoming part of our daily life.

The palm of the users plays an important role in the manipulation and detection of objects [1]. The interaction force between the objects and the palms determines the perception of contact, weight, shape, and orientation. At the same time, the friction forces produced by the object displacement can be perceived as friction forces produced by the slippage of the grasped objects.

Choi et al. [2], [3] introduced devices that deliver the sensation of weight and grasping of objects in VR successfully. These devices are located on the fingers, where unidirectional brakes create rigid grasping kinesthetic feedback. Tirado et al. [4] presented a device that delivers softness-hardness and stickiness feedback using electrical stimulation, touching a virtual object. Also, other devices were used to provide stimuli at the fingertips and potentially increase the immersion experience on VR [5]–[10]. Nevertheless, these devices deliver stimuli on the fingers, not on the palm.

The total number of tactile sensory units innervating the whole glabrous skin area of the human hand is around 17,023 *units* [11]. The density of the receptors located in

The authors are with Space Center, Skolkovo Institute of Science and Technology (Skoltech), 121205 Moscow, Russia {miguel.altamirano, jonathan.tirado, juan.heredia, D.Tsetserukou} @skoltech.ru

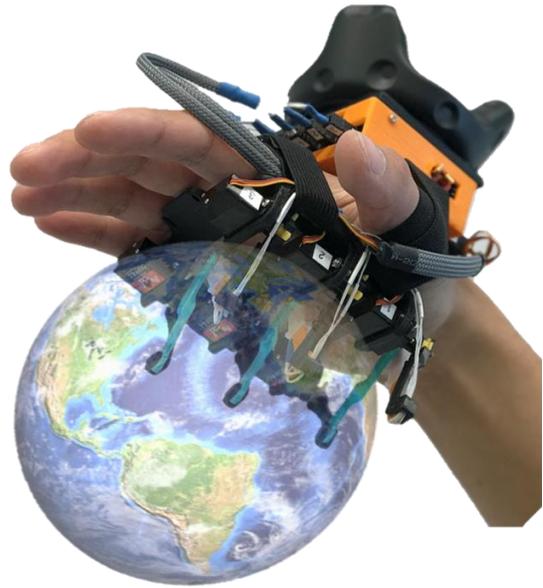

Fig. 1. A novel wearable haptic display $LinkGlide-S$.

the fingertips is five times bigger than the density in the palm. However, the overall receptor number on the palm is compensated by its largest area, having the 30% of all the SA-I receptors located at the hand glabrous skin. For this reason, it is essential to take advantage of the palm and develop devices that stimulate the most significant area by multi-contact points [12].

Kuchenbecker et al. [13] introduced a mechanical impedance control to render the forces of the wrist in a haptic device. Tsetserukou et al. [14] implemented a robust impedance control to realize safe and smooth human-robot interaction and to provide the effectiveness of the contact task performance. These papers show that impedance control can effectively modulate the sensation of stiffness. Tsykunov et al. [15] proposed a novel interaction strategy to guide the formation of quadrotors based on impedance control and vibrotactile feedback.

In the present work, we are introducing a novel wearable haptic interface $LinkGlide-S$ and an extended perception study of the multi-contact stimuli on the users' palm. A brief description of the preliminary design was presented in [16], and its integration with the VR systems [17]. The wearable tactile display $LinkGlide-S$ is designed to provide multi-contact and stimuli on the user's palm with two purposes: the realistic rendering of object perception during the detection and manipulation in VR environments and transferring the information during telemanipulation. This paper presents a novel wearable design for a hand-worn haptic display and introduces a multi-contact perception study on the palm.

We have integrated force sensors in each of the three contact points, allowing a force input in the system for the impedance control implementation. Additionally, the device was redesigned to improve its wearability and adapt to different hand shapes.

The first part of this work presents the $LinkGlide-S$ configuration. The second part illustrates the impedance control used to deliver the stiffness stimuli. In the last part, three experiments are carried out to evaluate the users' perception.

## II. LINKGLIDE-S WEARABLE HAPTIC DEVICE

$LinkGlide-S$ provides the sense of touch at three different points on the palm of the user, where multi-modal stimuli can potentially be delivered. The proposed device is based on the inverted five-bar linkages mechanism of M-shape introduced in [18]. The array of mechanisms generates the independent contact points covering the palm area. Each one of the contact points can be supplied by a vibration motor, and, potentially, by other stimulus generators, such as electro-stimulation. The force sensors at the end-effectors and the implementation of an impedance control provide the object softness sensation.

The $LinkGlide-S$ device was designed to adapt to the user's hand ergonomically, improving the operability of the fingers in comparison with the previous version. It has a total weight of 270 $g$ (89 $g$ for the VR tracker, and 181 $g$ for the device). These interface characteristics define a good wearability level according to [8].

Three 2-DoF inverted five-bar mechanisms, distributed in parallel planes, produce sliding force and multi-contact state at the palm by three contact points. Each of the systems has two PowerHD DSM44 servo motors, their angles of rotation are controlled to locate the single end-effector in the 2D coordinate system, as shown in Fig. 2b. Each end-effector creates a contact point at user's palm with different forces and glides on the palm shape applying an adaptive force.

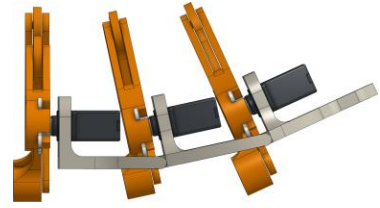

Fig. 3. Bending of the device using modular base. Lateral view of the device with a positive angle rotation.

The base of the device consists on three modules connected by metal pins, which allow the rotation of 15 deg. in both directions, to adapt to the position of the user's hand while opening and closing. The rotation of the device's base is shown in Fig. 3.

### A. Force Sensing

In each of the three contact points, one Force Sensitive Resistor (FSR) was installed, as shown in Fig. 4. The principal goal of the FSR is to measure the normal force generated in each contact point. The sensed force is an input for the impedance control, which is proposed to provide the sensation of the object stiffness. The different shapes and sizes of the user's palms are a challenge to overcome to deliver force and to ensure a contact point between the points $C_n(x_n, y_n)$ and the user's palm. The FSR data is used to solve this issue, monitoring if the contact is created at the right point, and if the correct force is delivered. The sensor is held by a component that was designed and 3D-printed. A semi-flexible material (TPU 95A, thermoplastic polyurethane) was used to print the FSR holder, which allows reducing the signal noise from the deformation of the sensor and creates a comfortable surface at the contact point. The holder is located on each end-effector with a groove where the FSR is inserted. Calibration is performed every time the device is active in a non-touch position.

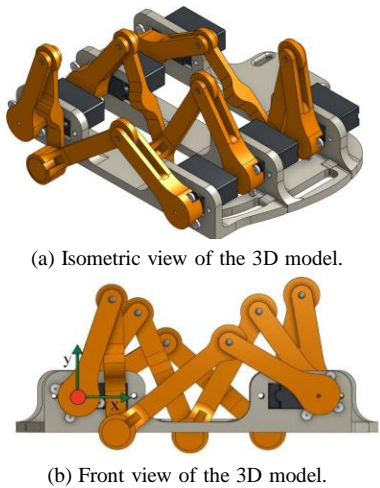

(a) Isometric view of the 3D model.

(b) Front view of the 3D model.

Fig. 2. 3D CAD model of wearable tactile display $LinkGlide-S$.

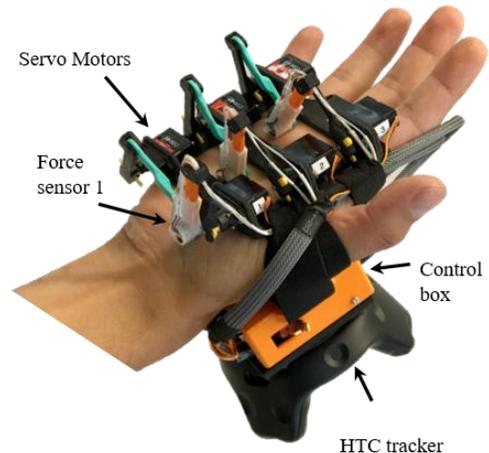

Fig. 4. $LinkGlide-S$ worn on the palm.

## B. Electronics and Software Design

The device is controlled using a Wi-Fi ESP8266 low-cost microchip. One Multiplexer is implemented to read the data from the three force sensors. A PWM Driver PCA9685 was used to control the 6 servo motors, it is connected to the microcontroller via I2C communication. One Li-Po 7.4 V battery is connected through a DC/DC converter to power the microcontroller and the servo driver.

A custom Python library was created to control the device. The device is connected to a server via Wi-Fi, from which it receives the coordinate of points $C_n(x_n, y_n)$ to reach in each interaction. The servomotors Power HD DSM44 allow a high speed of $0.07s/60°$. Therefore, the maximum reaction time, from when the command is called until the device reaches the further desired position, is of 0.15 s. There are, however, positions that can be reached on a lower time. The refresh frequency to the server is 100 [Hz].

## III. IMPEDANCE CONTROL CONFIGURATION

We propose an impedance control to implement rendering of softness by the $LinkGlide-S$ display between the virtual object and the hand. The stiffness on the contact point depends on the force delivered to the palm. A virtual mass-spring-damper link is introduced for the impedance model between each contact point and the palm of the user.

The impedance link is located in the axis normal to the palm surface ($y-axis$) of each contact point. The force delivered to the palm from the haptic device is generated in the virtual environment and represents the force with which the virtual object touches the virtual hand. The force sensors give feedback of the actual force applied to the hand, and the impedance control regulates this force. Therefore, the actual and virtual forces are the same.

When the virtual user approaches a virtual object, the impedance model is used to control the force to be delivered on the palm. The impedance control generates the end-effector trajectory that goes through the hand. Therefore, the device is in the same contact point but delivers more force to the user hand. The factors of the mass-spring-damping link can be modified to provide different stimuli.

In order to calculate the impedance correction term for the end-effector positions, we solve the second-order differential equation that represents the impedance model:

$$M_{di}\Delta\ddot{y} + D_{di}\Delta\dot{y} + K_{di}\Delta y = F_{ext}(t_i), \quad (1)$$

where $M_{di}$ is the desired mass of a virtual object in each point, $D_{di}$ is the desired damping, and $K_{di}$ is the desired stiffness. $\Delta y$ is the difference between the current position and the desired position of the contact point, and the $F_{ext}(t_i)$ is the sensed force. According to the selection of different parameters, the performance of the impedance model can render different stimuli on the user's palm. State space representation of (1) has the form:

$$\begin{vmatrix} \Delta y \\ \Delta \dot{y} \end{vmatrix} = A \begin{vmatrix} \Delta y \\ \Delta \dot{y} \end{vmatrix} + B F_{ext}, \quad (2)$$

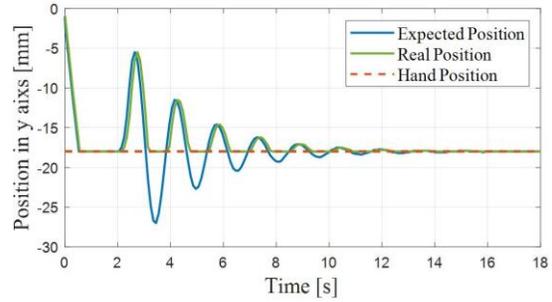

Fig. 5. Contact point expected position in $y-axis$ (blue line) and real position (green line) while under-damping response of the impedance control to simulate an object bouncing on the user's palm. The red line represents the user's palm position.

where $A = \begin{vmatrix} 0 & 1 \\ -\frac{K_d}{M_d} & -\frac{D_d}{M_d} \end{vmatrix}, B = \begin{vmatrix} 0 \\ \frac{1}{M_d} \end{vmatrix}$. In discrete timespace, after integration, we write the impedance equation as:

$$\begin{vmatrix} \Delta y_{k+1} \\ \Delta \dot{y}_{k+1} \end{vmatrix} = A_d \begin{vmatrix} \Delta y_k \\ \Delta \dot{y}_k \end{vmatrix} + B_d F_{ext}^k, \quad (3)$$

where $A_d = e^{AT}$, $B_b = (e^{AT} - I)A^{-1}B$, $T$ is the sampling time, $I$ is the identity matrix, and $e^{AT}$ is the state transition matrix. The matrix exponential is found from Cayley-Hamilton theorem, according to which every matrix satisfies its characteristic polynomial. Using these statements, we can find:

$$A_d = e^{\lambda T} \begin{bmatrix} (1-\lambda T) & T \\ -bT & (1-\lambda T -aT) \end{bmatrix} \quad (4)$$

$$B_d = -\frac{c}{b} \begin{bmatrix} e^{\lambda T}(1-\lambda T -1) \\ -bTe^{\lambda T} \end{bmatrix}, \quad (5)$$

where $\lambda$ is the eigenvalue variable of the matrix $A$, $a = -\frac{D_d}{M_d}$, $b = -\frac{K_d}{M_d}$, and $c = \frac{1}{M_d}$. $A_d$ and $B_d$ matrices can be used to calculate the current $y_{imp}^c$ position of the impedance model using equation (5).

The implementation of the impedance control is done in one dimension, changing the position in the $y-axis$. The force from each contact point is constantly used to calculate the independent impedance of each contact point. In Fig. 5, the contact point expected position, real position, and hand position are rendered while the impedance control is activated using an under-damped response.

## IV. EXPERIMENTAL EVALUATION

Three different experiments were carried out: $i$) to evaluate the user perception of patterns delivery at the human palm, $ii$) to assess how the device helps to improve the user experience in a Virtual Reality environment, and $iii$) to examine the user perception of stiffness patterns using impedance control.

### A. Experiment 1: Static Pattern Recognition

The experiment focuses on the perception of patterns on the palm using the $LinkGlide-S$ interface.

**Participants:** Sixteen volunteers, six women and ten men (23.56±2.22 years old) participated and completed the test. The participants were informed about the experiment and filled out the consent form. This study was approved by the local Institutional Review Board with reference number "MoM Protocol No.5 dd. July 21, 2021". None of them reported any deficiencies in sensorimotor function, and all of them were right-handed.

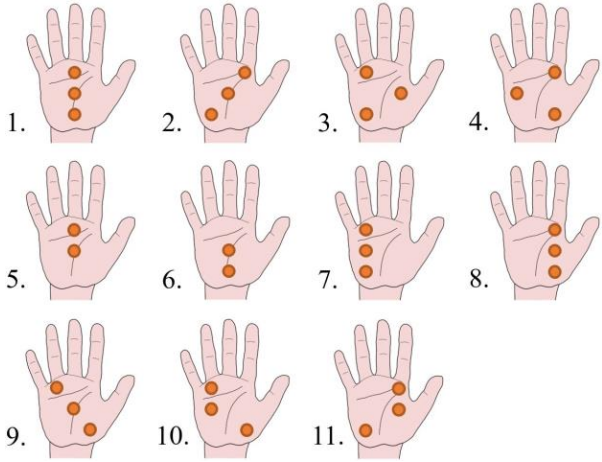

Fig. 6. The designed tactile patterns for the static position experiment.

TABLE I
CONFUSION MATRIX FOR ACTUAL AND PERCEIVED STATIC PATTERNS ACROSS ALL SUBJECTS

| % | | Answers (predicted class) | | | | | | | | | | |
|---|---|---|---|---|---|---|---|---|---|---|---|---|
| | | 1 | 2 | 3 | 4 | 5 | 6 | 7 | 8 | 9 | 10 | 11 |
| Patterns (actual class) | 1 | 86.3 | 5.0 | 0.0 | 1.3 | 1.3 | 5.0 | 1.3 | 0.0 | 0.0 | 0.0 | 0.0 |
| | 2 | 0.0 | 62.5 | 7.5 | 1.3 | 0.0 | 0.0 | 0.0 | 0.0 | 2.5 | 0.0 | 26.3 |
| | 3 | 0.0 | 0.0 | 88.8 | 0.0 | 0.0 | 0.0 | 0.0 | 0.0 | 0.0 | 1.3 | 10.0 |
| | 4 | 1.3 | 3.8 | 1.3 | 88.8 | 0.0 | 0.0 | 0.0 | 0.0 | 0.0 | 1.3 | 3.8 |
| | 5 | 5.0 | 0.0 | 0.0 | 0.0 | 80.0 | 15.0 | 0.0 | 0.0 | 0.0 | 0.0 | 0.0 |
| | 6 | 6.3 | 0.0 | 0.0 | 0.0 | 1.3 | 91.3 | 0.0 | 1.3 | 0.0 | 0.0 | 0.0 |
| | 7 | 0.0 | 5.0 | 8.8 | 1.3 | 1.3 | 1.3 | 71.3 | 0.0 | 11.3 | 0.0 | 0.0 |
| | 8 | 1.3 | 2.5 | 0.0 | 6.3 | 2.5 | 0.0 | 1.3 | 83.8 | 0.0 | 0.0 | 2.5 |
| | 9 | 10.0 | 0.0 | 1.3 | 2.5 | 0.0 | 2.5 | 2.5 | 5.0 | 65.0 | 10.0 | 1.3 |
| | 10 | 1.3 | 1.3 | 0.0 | 7.5 | 1.3 | 0.0 | 0.0 | 0.0 | 16.3 | 71.3 | 1.3 |
| | 11 | 0.0 | 25.0 | 10.0 | 0.0 | 0.0 | 0.0 | 0.0 | 0.0 | 0.0 | 0.0 | 65.0 |

**Experimental setup:** The subject was asked to sit in front of a desk and to wear the $LinkGlide-S$ device on the right hand, which was located palm up with the back resting on the table. To reduce the external stimuli, we asked the users to use headphones with white noise. A physical barrier obscured the visibility of the hand. During the experiment, the patterns were displayed on a screen visible to the participants.

**Method:** In this experiment, eleven tactile patterns have been designed, as shown in Fig. 6, where the red circles represent the contact points on the palm.

Each pattern was delivered for 2 seconds on the user's palm. Each pattern was displayed five times in random order.

**Results and discussion:** The experimental results are summarized in a confusion matrix (see Table I).

To determine the statistical differences in pattern recognition, we conducted one-factor ANOVA without replication with a chosen significance level of $\alpha < 0.05$. The sphericity and normality assumptions were examined and no violations were detected. According to the ANOVA results, there is a statistical significant difference in the pattern recognition of the users, $F(10, 165) = 2.6724$, $p = 0.0047$, $\eta_p^2 = 0.1393$, 90%, suggesting a large effect on the recognition due to the different patterns rendered.

The paired t-tests with one-step Bonferroni correction showed statistically significant differences between the recognition of pattern 4 and pattern 11 with a chosen significance level of $\alpha < 0.15$ ($p = 0.1324 < 0.1$, $Hedges g = -1.14$), and between pattern 6 and pattern 11 ($p = 0.1220 < 0.1$, $Hedges g = -1.15$). The open-source statistical package Pingouin for Python was used for the statistical analysis.

The average recognition rate is 78.1%. Patterns 2 and 9 have lower recognition rates of 62.5% and 65%, respectively. The two patterns with lower recognition have similar designs, both diagonal, but in different directions. We can observe that, in both cases, the patterns are confused with the patterns 11 and 10, respectively. Those patterns have the middle point in a different position. This effect shows that the central part of the palm has to be stimulated more accurately because of its lower contact during manipulation.

### B. Experiment 2: Interaction with Objects in VR

This experiment aims to investigate the realistic perception of object interaction using the proposed device in VR. Three patterns from experiment 1 were selected and used to render the spherical object surface. The results will help to develop the best strategy to create tactile sensation with the available three contact points.

**Participants:** Seven participants, two women and five men (23.86±2.73 years old), completed the experiment. None of them reported any deficiencies in sensorimotor function, and all of them were right-handed.

**Experimental Setup:** The users were asked to wear the device on the palm and a head-mounted display (HMD) HTC Vive Pro. A VR environment was designed in the game engine Unity 3D. The 3D environment consists of three spheres of 25 $cm$ in diameter located at 1.2 $m$ floating over the virtual floor, as shown in Fig. 7.

**Method:** The users were asked to grade the similarity of the haptic sensation rendered by the device patterns to the real interaction with the objects. In the virtual 3D environment, the user was asked to interact for one minute with each of three spherical objects. When the users' virtual hand contacts the surface of the spherical objects the specific tactile pattern is activated by the haptic display. Two of the patterns were selected according to the high recognition rate from the experiment 1, i.e. patterns 1 and 3. Additionally, we opted for one diagonal pattern 2 arguing that that there will be no ambiguity with the symmetrical one anymore. The selected tactile patterns are patterns 1, 2, and 3 from Fig. 6.

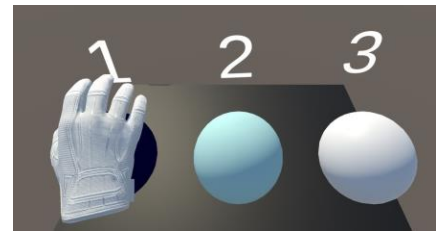

Fig. 7. Virtual Reality environment for the experiment 2.

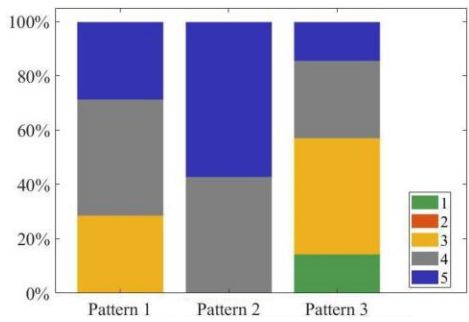

Fig. 8. The similarity perception rate of the delivered patterns, where 1 is the lowest and 5 is the highest similarity level during the interaction with the three spherical object surfaces.

TABLE II
CONFUSION MATRIX FOR IMPEDANCE PATTERNS

| % | | Answers (predicted class) | | |
|---|---|---|---|---|
| | | 1 | 2 | 3 |
| Patterns (actual class) | 1 | 86.7 | 13.3 | 0.0 |
| | 2 | 16.7 | 76.7 | 6.7 |
| | 3 | 0.0 | 10.0 | 90.0 |

TABLE III
CONFUSION MATRIX FOR FORCE CONTROL

| % | | Answers (predicted class) | | |
|---|---|---|---|---|
| | | 4 | 5 | 6 |
| Patterns (actual class) | 4 | 86.7 | 13.3 | 0.0 |
| | 5 | 10.0 | 86.7 | 3.3 |
| | 6 | 0.0 | 13.3 | 86.7 |

The participants interacted with each virtual spherical object for one minute. The interaction can be done from any part of the sphere. The participants graded the similarity, from 1 (lowest) to 5 (highest), to the real interaction with a spherical object.

**Results and discussion:** The similarity perception rates during the interaction with the spherical object surfaces are presented in Fig. 8.

To determine the statistical difference between the users' grades, Friedman test with a chosen significance level of $\alpha < 0.05$ was performed. Statistically significant difference was found between the similarity patterns rates ($p = 0.0067 < 0.05$). Wilcoxon tests were done for pairwise comparison for the three patterns' similarity perception rates. According to the statistical significance difference, it was found that interaction with a 3D sphere with pattern 2 is more realistic comparing with pattern 3 ($p = 0.0235$) and pattern 1 ($p = 0.0455$). Based on this evidence, the pattern with the lower recognition rate in the first experiment obtained the highest similarity perception during the interaction with the spherical object surfaces in the VR environment.

*C. Experiment 3: Stiffness Representation by Impedance Control*

This evaluation was performed to analyze the human perception of the different stiffness of objects delivered on the palm by the haptic interface. The information from the force sensors in the contact points was used to deploy the control techniques. We propose two control approaches to tactile interaction:

- Force Control: the device end-effectors move to the palm until one of it measures the limit force. Then, the device returns to a no-contact position.
- Impedance Control: the device interacts with the user's palm using the impedance control equations described in section III.

For both cases, the three contact points are located using pattern 2 of the previous experiments. Since pattern 2 of the previous experiment had the highest recognition rate, we consider it as the best one to test the control techniques.

We tested three different stiffness values in this experiment: strong, medium, and low, for each control technique. The parameters for the tactile patterns are as follows: Pattern 1 (strong stiffness, impedance control): inertial coefficient of 1.2 $kg$, a stiffness coefficient of 20 $Ns/m$, and a damping coefficient of 1 $N/m$. Pattern 2 (medium stiffness, impedance control): inertial coefficient of 0.6 $kg$, a stiffness coefficient of 3 $Ns/m$, and a damping coefficient of 1 $N/m$. Pattern 3 (low stiffness, impedance control): inertial coefficient of $0.6 kg$, a stiffness coefficient of 1 $Ns/m$, and a damping coefficient of 1 $N/m$. Pattern 4 (strong stiffness, force control): limit force of 4 $N$. Pattern 5 (medium stiffness, force control): limit force 2.5 $N$. Pattern 6 (low stiffness, force control): limit force 1 $N$.

**Participants:** Six participants completed the experiments, two women and four men (24.7 ±3.4 years old). None of them reported any deficiencies in sensorimotor function, and all of them were right-handed.

**Experimental Setup:** The user was asked to sit in front of a desk and to wear the $LinkGlide-S$ device on the palm. To reduce the external stimuli, the users wore headphones with white noise. A physical barrier interrupted the vision of the users to their hand. During the experiment, the patterns were displayed on a screen.

**Method:** Before each section of the experiment, a training session was conducted, where all the tactile patterns were delivered two times to get acquainted with the stimuli. Each pattern was delivered on their palm five times in a random order. The user was asked to specify the force intensity that corresponds to the delivered pattern.

This study was divided into two parts. During the first part, the experimental procedure was implemented for the tactile patterns using impedance control. In the second part, the experimental procedure was implemented for the tactile patterns using force control.

**Results and discussion:** The experimental results are summarized in two confusion matrices for actual and perceived patterns across all subjects. Table II contains the data from the impedance control patterns and Table III shows the data from the force control patterns.

The patterns perception was analyzed using Friedman test with a chosen significance level of $\alpha < 0.05$. Statistically significant difference was found between the patterns delivered by the impedance control ($p = 0.0421 < 0.05$). No statistically significant difference was found between the patterns delivered by the force control ($p = 0.6065 > 0.05$). Wilcoxon tests were performed for pairwise comparison of the three patterns that implemented the impedance control. According to the statistical significance difference, pattern 3 is more recognizable than pattern 2 ($p = 0.034 < 0.05$).

The difference in the results indicates that users can perceive and distinguish different information by modifying the impedance parameters. The impedance control delivers a smooth force contact to the user's hand. We can conclude that the impedance control effectively presents the stiffness of objects to the user's palm.

## V. Conclusions and Future Work

In this work, a novel wearable multi-contact tactile display was presented. An extended perception study of the multi-contact stimuli was performed. The implementation of the impedance control for object stiffness rendering was introduced and analyzed.

Three different experiments were carried out. The first experiment evaluated the user perception of the tactile patterns at the users' palms. The second study analyzed the users' sensations while interacting with spherical object surfaces in VR environment, using selected tactile patterns from the experiment 1. The last experiment examined the user perception of stiffness patterns using impedance control.

In experiment 1, the average recognition rate was 78.1%. Five patterns were perceived with a recognition rate of 80% or over. $LinkGlide-S$ can potentially be used in HRI applications, in future work this application will be explored. In experiment 2, three patterns from experiment 1 were selected to the perception of interaction with three spheres in VR. The results showed that the tactile pattern 2 delivered the most realistic sensation in comparison with the other patterns. In experiment 3, a perception analysis between force and impedance control for pattern recognition was perfomed. The results indicate that users can perceive and distinguish different patterns by modifying the impedance parameters.

In future work, different impedance parameters and tactile patterns will be explored in VR applications to enhance the object perception for a variety of shapes. The collaborative robot teleoperation will be implemented, and the users' perceptions will be explored, such as the force perception from a remote robotic gripper during the grasping of deformable objects.

$LinkGlide-S$ can be used to represent the sensation of virtual objects on the palm and, potentially, at other places of the body (the structure of the device can be adaptable to any body surface). The proposed display can be used in affective haptics (e.g. handshake presentation), robot bilateral teleoperation, and teaching robotic systems using the human skills, improving the user immersion.

## Acknowledgements

The reported study was funded by CNRS and RFBR according to the research project No. 21-58-15006.